\documentclass{article}

\usepackage{amsmath, amssymb}

\newtheorem{theorem}{Theorem}[section]
\newtheorem{lemma}[theorem]{Lemma}
\newtheorem{proposition}[theorem]{Proposition}

\newtheorem{definition}[theorem]{Definition}

\newtheorem{remark}[theorem]{Remark}

\newcommand{\mcA}{{\mathcal{A}}}
\newcommand{\mcM}{{\mathcal{M}}}
\newcommand{\beq}{\begin{eqnarray}}
\newcommand{\eeq}{\end{eqnarray}}

\newcommand{\Msec}{{\mcM}^{(2)}}
\newcommand{\db}{{\tilde{d}}^2}
\newcommand{\mE}{\mathcal{E}}
\begin{document}
\title{Construction of graded differential algebra\\ with ternary differential}
\author{Nadezda Bazunova \thanks{The author was supported  by ETF Grant
5634.}}
\date{}

\maketitle

\begin{abstract}
In this article, we describe the construction of graded
$q$-differential algebra with ternary differential satisfying the
property $d^3=0$ and the $q$-Leibniz rule. Our starting point is
coordinate first order differential calculus on some complex algebra
$\mcA$ and the corresponding bimodule of second order differentials.
\end{abstract}

\section*{Introduction}

The generalization of classical differential calculus for the
differential $d^N=0$ and $q$-deformed Leibniz rule
\beq
 \label{L-rule}
 d(\omega \theta)= (d\omega )\theta + q^{\bar{\omega}}\omega (d\theta),
  \eeq
has been
worked out and studied in \cite{DV-K}, and specifically for the
ternary case $d^3=0$ in \cite{RK,A-B}. In such generalizations,
the condition $d^2\neq 0$ implies the appearing of higher order
differentials of variables. The differential calculus with
differential $d^3=0$ on an associative algebra with $n$ generators
and quadratic relations, and on the quantum plane specifically,
was constructed in \cite{BBK}. Also, such a construction was
realized on superspaces with one and two parametric quantum groups
as symmetry groups in \cite{B-H-H-Z}.

In this paper, we consider the construction of graded
$q$-differential algebra with ternary differential on an arbitrary
complex associative unital algebra. That is, we extend the first
order differential calculus to a differential $d$ satisfying  the
property $d^3=0$ and the $q$-Leibniz rule (\ref{L-rule}), where the
complex number $q$ is a third root of unity \cite{DV-K}.

Our inspiration is twofold. On one hand, we have the approach of S.
Woronowicz \cite{W} for constructing a quantum de Rham complex, for
an ordinary differential ($d^2=0$), starting from some first order
(coordinate) differential calculus (see \cite{B-Kh}).
On the other hand, we generalize the method applied in
\cite{DV-K,DV-K2} for the universal differential calculus and
$N$-ary differential.

We shall use the notation $[n]_q=1+q+q^2+\dots +q^{n-1}$.  In our
case, $q=e^{i\frac{2\pi}{3}}$, we have $[3]_q=1+q+q^2=0$.

On the other hand, the $N$-differential graded algebras (for $N>2$)
with the usual Leibniz rule
$$
d(\omega \theta)= (d\omega )\theta + (-1)^{\bar{\omega}}\omega
(d\theta)
$$
also are of interest. In the recently appeared paper \cite{A-D},
authors examine the construction of an $N$-differential graded
algebra with the usual Leibniz rule and moduli space of deformations
of the differential of an $N$-differential graded algebras.

\section{First order differential calculus with right partial derivatives}

Let us  remind the construction of first order coordinate
differential calculus considered in \cite{B-Kh} in detail.

Let $\mcA$ be an algebra, $\mcM$ be an $\mcA$-bimodule, $d^1:\mcA
\rightarrow \mcM$ be a linear mapping such that the Leibniz rule
 \beq
 \label{L-rule1}
d^1(uv)=d^1(u)v+ud^1(v), \quad \forall\ u,\,v\in\mcA.
  \eeq
is fulfilled.

 \begin{definition}
 {The triple $\{ \mcM, \ \mcA,\ d^1 \}$ is called
a first order differential calculus (FODC) on an algebra $\mcA$ with
values in $\mcM$. Elements of $\mcM$ are called one-forms, while
$d^1$ is called differential.} \end{definition}

We will consider the FODC for the case that $\mcA$ is a complex
associative unital free algebra with generators $\{x^1,\ x^2, \dots
, x^n\}\subset \mcA$ and $\mcM$ is a free right module with basis
$\{dx^1,\ dx^2, \dots ,\ dx^n\}\subset \mcM ,\ n\in \mathbb{N}$.

A structure of bimodule on the free right module $\mcM$ can then be
defined by means of an algebra homomorphism
$\xi=(\xi^j_k):\mcA\rightarrow {\rm Mat}_{n}\mcA$ such that
 \beq
  u\,dx^j=dx^k\xi(u)^j_k, \quad \forall u\in\mcA,\ j,\,k=1,\dots,\,n .\nonumber
  \eeq

As is well known, the first order differential of an arbitrary
element $v\in\mcA$ can be defined by the decomposition
 \beq
 d^1(v)=dx^kD_k(v),\quad \forall\ v\in\mcA,\nonumber
  \eeq
where $d^1(x^i)=dx^i$, and the linear maps $D_k\colon
\mcA\rightarrow\mcA$ are called right partial derivatives. It is
also assumed that $D_k(x^i)= \delta^i_k$, i.e.\ we are  dealing with
the so-called coordinate differential.

Then,  as it follows from the Leibniz rule (\ref{L-rule1}), the
partial derivatives $D_k$ and the homomorphism $\xi$ are connected
by the formula
 \beq
 D_k(uv)=D_k(u)v+\xi(u)^j_k D_j(v), \quad \forall\ u,\, v\in\mcA.\nonumber
  \eeq

We would like to emphasize that the property
$d^1\mcA\,\mcA=\mcA\,d^1\mcA$ follows immediately from the Leibniz
rule (\ref{L-rule1}). We also assume that $\mcM=d^1\mcA\,\mcA$. This
means that the $\mcA$-bimodule $\mcM$ doesn't contain extra elements
besides  differentials.

\section{Graded differential algebra with $d^3=0$}

At the beginning of this section, we remind the definition of graded
$q$-differential algebra with ternary differential $d^3=0$ proposed
in \cite{DV-K,DV-K2}, where such graded (universal) $q$-differential
algebra with differential $d^N=0$ has been constructed starting from
the universal first order differential calculus corresponding to a
given algebra $\mcA$.
 \begin{definition}{The pair $\{ \Omega, d\}$, where
$\Omega=\oplus_{n\geq 0}\,\Omega^{(n)}$} is a graded algebra over
the ring $\mcA=\Omega^0$, and $d\colon\,\Omega\rightarrow \Omega,\
d=\oplus_{n\geq 1}\,d^n$ is a homogeneous linear map of grade one,
where $d^{n+1}\colon\,\Omega^{(n)}\rightarrow\Omega^{(n+1)}$, such
that the $q$-Leibniz rule
 \beq
 \label{L-rule2}
d(\omega \theta)= (d\omega )\theta + q^{n}\omega (d\theta),\quad
\textrm{for}\ \omega \in \Omega^{(n)},\ \theta \in \Omega,
 \eeq
and the property $d^3=0$ are satisfied, is called a graded
differential algebra with ternary differential.\end{definition}

The elements of $\Omega^{(n)}$ are referred to as $n$-forms.
\begin{remark}
The triple $\{\mcM ,\ \mcA, d^1\}$ is a trivial example of graded
differential algebra with ternary differential $\{\Omega,\, d\}$,
with $\Omega^{(0)}=\mcA$,\, $\Omega^{(1)}=\mcA$,\, $\Omega^{(n)}=0$
for $n>1$, where the differential $d$ coincides with $d^1$ on the
algebra $\mcA$.
\end{remark}

Before specification of such an algebra in the case of coordinate
calculus with right partial derivatives, we will define the second
order differential (SOD).

Let $\Msec$ be a free  right module over $\mcA$ generated by
elements $\{d^2x^1,\ d^2x^2, \dots ,$ $\ d^2x^n\}\subset \Msec ,\
n\in \mathbb{N}$, and let us assume that the structure of bimodule
on it is set  by means of the same homomorphism
$\xi=(\xi^j_k):\mcA\rightarrow {\rm Mat}_{n}\mcA$ as in the case of
the bimodule $\mcM$, i.e.
 \beq
 \label{ud2}
  u\,d^2x^j=d^2x^k\xi(u)^j_k, \quad \forall\ u\in\mcA ,\ j,\,k=1,\dots ,\,n .
  \nonumber
  \eeq

Define a linear mapping $\db\colon\,\mcA\rightarrow\Msec$, such that
 \beq
\db(x^i)=d^2x^i,\qquad \db(u)=d^2x^iD_i(u), \nonumber
 \eeq
where $D_i$ are the same right partial derivatives as in the case of
FODC above. As a consequence  the Leibniz rule
 \beq
 \label{L2(1)-rule}
\db(uv)=\db(u)v+u\db(v),\quad \forall\; u,\;v\in\mcA, \nonumber
 \eeq
is automatically satisfied. Let us remark that, because of the
obvious correspondence  $dx^i\leftrightarrows d^2x^i$, the
 bimodules $\mcM$ and $\Msec$ are isomorphic. We
also assume that the elements $d^2x^i$ have the grade 2, and refer
to elements of $\Msec$ as 2-forms.

We call the triple $\{\Msec,\;\mcA,\;\db\}$ the canonical second
order differential (SOD) associated to a given FODC $\{ \mcM, \
\mcA,\ d^1 \}$ (see \cite{BBK} for non-canonical SOD).

Let us define a graded $\mcA$-bimodule $\mE=\mcM^{(1)}\oplus\Msec$,
where $\mcM^{(1)}\equiv\mcM$. Construct the tensor algebra of $\mE$
over the ring $\mcA$:
 \beq
 T_{\mcA}\mE=\mcA\oplus\mE\oplus\mE^{\otimes 2}\oplus\mE^{\otimes
 3}\oplus\dots =\bigoplus_{n\geq 0}\mE^{\otimes n}\nonumber
 \eeq
 where $\mE^{\otimes n}$ means $\mE\otimes_{\mcA}\dots\otimes_{\mcA}\mE$.

 After expanding and rearranging terms in $T_{\mcA}\mE$, we obtain
 an $\mathbb{N}$-graded algebra
 \beq \label{Talg}
 T_{\mcA}\mE=\bigoplus_{n\geq 0}T^n,
 \eeq
 with a unique graduation compatible with graduation of $\mE$ (see \cite{DV-K2}).
 Here $T^0=\mcA,\ T^1=\mcM\equiv\mcM^{(1)},\
T^2=\Msec\oplus\mcM^{(1)}\otimes_{\mcA}\mcM^{(1)}$, and
for $n>2$ \beq\label{Tn} T^{n}=\bigoplus_{\substack{m\leq n,\, a_i=1,2;\\
a_1+\dots+a_m=n}
}\mcM^{(a_1)}\otimes_{\mcA}\dots\otimes_{\mcA}\mcM^{(a_m)} \eeq
consists all possible tensor products of total grade equal to $n$.
In the sequel, when writing $\omega \otimes \theta $ for elements
from $T_{\mcA}\mE$, we mean tensor product over the algebra $\mcA$,
that is $\omega \otimes_{\mcA} \theta \doteq \omega \otimes \theta
$.

Due to the fact that the tensor product in (\ref{Tn}) is taken over
the ring $\mcA$, any homogenous element of grade $n$ has a unique
decomposition in the form
 \beq
 \label{decom}
d^{a_1}x^{i_1}\otimes\dots\otimes d^{a_m}x^{i_m}r_{i_1\dots i_m}
 \eeq
where $m\leq n,\,\, r_{i_1\dots i_m}\in\mcA$ and the sum in
(\ref{decom}) is over all possible  choices of $a_1,\dots, a_m$,
where $a_i\,{\in}\,\{1,2\}$, such that $a_1+\dots+a_m=n$.

Now we are in a position to define a grade one linear mapping
$d:T_{\mcA}\mE\rightarrow T_{\mcA}\mE$ by:
 \beq
 \label{d}
 \begin{array}{l}
 d( d^{a_1}x^{i_1}\otimes\dots\otimes d^{a_m}x^{i_m}r_{i_1\dots i_m})=
 \sum^m_{j=1}q^{(a_1+\dots+a_{j-1})}
 \end{array}\qquad\qquad\qquad\\[2pt]
 \begin{array}{r}
 d^{a_1}x^{i_1}\otimes\dots\otimes
 d^{a_{j-1}}x^{i_{j-1}}\otimes
d^{a_j+1}x^{i_j}\otimes d^{a_{j+1}}x^{i_{j+1}}
 \dots\otimes d^{a_m}x^{i_m}r_{i_1\dots i_m}\quad
\\[4pt]
+\,q^{(a_1+\dots+a_{m})} d^{a_1}x^{i_1}\otimes\dots\otimes
d^{a_m}x^{i_m}\otimes dx^sD_s(r_{i_1\dots i_m})\nonumber
\end{array}
 \eeq
and $d^3x^i=0$ is assumed. One already sees that $d(x^i)=dx^i$,
$d^2(x^i)=d(dx^i)=d^2x^i$ and $d^3x^i=d^2(dx^i)=d(d^2x^i)=0$, which
means that $d$  is a prolongation of $d^1$ and $\db$. Nevertheless,
$\db$ and $d^2$ are not identical, since
 \beq
d^2u=d(dx^iD_i(u))=\db u+qdx^i\otimes dx^jD_j(D_i(u))\,.\nonumber
 \eeq
Similarly, a simple calculation shows that
 \beq
 \begin{array}{l}
d^3u=d(\db u+qdx^i\otimes dx^jD_j(D_i(u)))\nonumber
\\[4pt]
\quad\,\ \ = q[2]_q\,d^2x^i\otimes dx^jD_j(D_i(u))\nonumber
\\[4pt]
\qquad \qquad  +\,q^2dx^i\otimes d^2x^jD_j(D_i(u)) +dx^i\otimes
dx^j\otimes dx^kD_k(D_j(D_i(u)))\nonumber
 \end{array}
 \eeq
does not vanish in general. It should also be noticed that the
definition (\ref{d}) agrees with the Leibniz rule (\ref{L-rule2}).
However, for elements which are  not written in the canonical form
(\ref{decom}), the Leibniz rule does not hold in general. Because of
these, the tensor algebra $T_{\mcA}\mE$ equipped with the operator
$d$, as defined above, is not a graded differential algebra with
ternary differential.

Our aim now is to construct a minimal graded ideal
$I_q=\bigoplus_{n\geq 0} I^{(n)}_q$ in $T_{\mcA}\mE$, such that
$dI_q\subseteq I_q$. We call such an ideal $d$-compatible. Here
grading is meant with respect to $T_{\mcA}\mE$: $I_q^{n}\subset
T^n$.

In the construction of the ideal $I_q$, we shall follow a general
technique developed in \cite{B-Kh}. The basic relations are
 \beq
 \label{basic}
 x^idx^j-dx^k\xi_k^{ij}=0\quad \ {\rm and}\quad \
x^id^2x^j-d^2x^k\xi_k^{ij}=0\eeq where
$\xi_k^{ij}\doteq(\xi(x^i))_k^{j}\in\mcA$. Consecutive
differentiation of (\ref{basic}) by means of (\ref{L-rule2}) leads
to the following set of homogeneous elements in $T_{\mcA}\mE$:
 \beq
  \label{rel1}
  & dx^i\otimes dx^j-qdx^k\otimes d\xi_k^{ij},&\\[4pt]
  \label{rel-2-2}
 & dx^i\otimes d^2x^j-q^2d^2x^k\otimes d\xi_k^{ij},&\\[4pt]
 \label{rel2}
 & d^2x^i\otimes dx^j+(1-q)d^2x^k\otimes d\xi_k^{ij}-
  q^2dx^k\otimes d^2\xi_k^{ij},&\\[4pt]
 \label{rel3}&
d^3\xi_k^{ij},&\\[4pt]
\label{rel4}& d^2x^i\otimes d^2x^j-qd^2x^k\otimes d^2\xi_k^{ij} ,&
  \eeq
  which, in fact, are generating relations for $I_q$. Here
 \beq
  \begin{array}{ll}
  d\xi_k^{ij}=dx^lD_l(\xi_k^{ij});&\nonumber\\[6pt]
  d^2\xi_k^{ij}=d^2x^lD_l(\xi_k^{ij})+qdx^l\otimes
  dx^mD_m(D_l(\xi_k^{ij}));&\nonumber\\[6pt]
d^3\xi_k^{ij}=q[2]_q\,d^2x^l\otimes dx^mD_m(D_l(\xi_k^{ij}))+q^2dx^l
\otimes d^2x^mD_m(D_l(\xi_k^{ij}))&\nonumber\\[4pt]
\hspace{5.cm} +\,dx^l\otimes dx^m\otimes
dx^pD_p(D_m(D_l(\xi_k^{ij}))).\nonumber
 \end{array}
\eeq

The main result of our paper is the following
\begin{theorem} \label{mainthm}
Let $\Omega(\mcA, \mcM)$ be the quotient algebra
 \beq
 \label{GDA}
 \Omega(\mcA,\mcM)\doteq\frac{T_{\mcA}\mE}{I_q}=\bigoplus_{n\geq
0}\frac{T^n}{I^n_q}\doteq\bigoplus_{n\geq 0}\Omega^{(n)}(\mcA,
\mcM),
 \eeq
 where the homogeneous ideal $I_q \subset T_{\mcA}\mE$ is generated
by the set of elements (\ref{basic})\,--\,(\ref{rel4}), and the
operator \mbox{$d\colon\,\Omega(\mcA, \mcM)\rightarrow \Omega(\mcA,
\mcM)$} is defined by the formula (\ref{d}).\par
 Then the pair $\{\Omega(\mcA,
\mcM), d\}$ is an $\mathbb{N}$-graded differential algebra with
ternary differential, that is the properties $d^3=0$ and $q$-Leibniz
rule:
 \beq
d(\omega \theta)= (d\omega)\,\theta+q^n\omega\,(d\theta),\quad
\textrm{where}\quad \omega\in\Omega^{n}(\mcA,\,\mcM),\quad
\theta\in\Omega(\mcA, \mcM),\nonumber
 \eeq
are satisfied.
\end{theorem}
In order to prove this theorem, we need additional results, which we
formulate in Proposition \ref{prop}. We establish it and the proof
of theorem \ref{mainthm} in Appendix A and in Appendix B,
respectively.

 To this end, remark that, for example, the bimodule $\Omega^{(2)}(\mcA,\mcM)$
has the structure
 \beq
\Omega^{(2)}(\mcA,\mcM)=\frac{\mcM\otimes\mcM\oplus\Msec}{I_q^{(2)}},\nonumber
 \eeq
where $I_q^{(2)}$ denotes the sub-bimodule in $T^2$ generated by
relations (\ref{rel1}). Now\\ \mbox{
$d^2\colon\,\mcA\rightarrow\Omega^{(2)}(\mcA,\mcM)$} satisfies the
q-Leibniz rule
 \beq
 d^2(uv)=d^2(u)v+{[2]}_q\;
 du\;dv+ud^2v, \quad \forall \ u,\,v\in\mcA,\nonumber
 \eeq
as expected \cite{DV-K}.

\appendix
\section{}
\begin{proposition}\label{prop}
Let the ideal $I_q \subset T_{\mcA} \mE$ be generated by the
elements\par {$
 x^idx^j-dx^k\xi_k^{ij}$,
$\ x^id^2x^j-d^2x^k\xi_k^{ij}
 $,
 and (\ref{rel1})\,--\,(\ref{rel4}).}\par
  Then in the tensor algebra $T_{\mcA} \mE$  equipped with the
operator $d$, the relations
  \beq
  \label{rel12}
  & dv\otimes dx^j=qdx^k\otimes d\xi (v)_k^{j}\quad (mod\, I_q),&\\[4pt]
  \label{rel-2-22}
 & dv\otimes d^2x^j=q^2d^2x^k\otimes d\xi (v)_k^{j}\quad (mod\, I_q),&\\[4pt]
 \label{rel22}
 & d^2v\otimes dx^j=(q-1)d^2x^k\otimes d\xi (v)_k^{j}+
  q^2dx^k\otimes d^2\xi (v)_k^{j}\quad (mod\, I_q),&\\[4pt]
 \label{rel32}&
 d^3\xi (v)_k^{j}=0\quad (mod\, I_q),&\\[4pt]
 \label{rel42}&
 d^2v\otimes d^2x^j=qd^2x^k\otimes d^2\xi (v)_k^{j}\quad (mod\, I_q)&
  \eeq
are true for all $v\in\mcA$.
\end{proposition}
 \textbf{Proof.\quad } In case of $v=x^i, \ i=1,\dots ,\,n$, we have
relations directly obtained from the generators of the ideal $I_q$
(\ref{rel1})\,--\,(\ref{rel4}). This enables us to do the proof of
the proposition by induction.

 Assuming (\ref{rel12})\,--\,(\ref{rel42}) hold for
 $v_i \in \mcA$, we will prove their validity for $v=x^i\, v_i$.
 Combining the basic relations (\ref{basic}), our assumptions,
 and the definition of the operator $d$, we obtain the following equalities:
 \beq
 \begin{array}{rl}
dv\otimes dx^j=&
%
dx^i\otimes dx^k\,\xi(v_i)_k^{j}+ q\,x^i\,dx^k\otimes
d\xi(v_i)_k^j\nonumber\\[3pt]
=&q\,dx^l\otimes d\xi_l^{i k}\,\xi(v_i)_k^{j}+q\,dx^l\xi_l^{i
k}\otimes d\xi(v_i)_k^j
%
=qdx^l\otimes d\xi(v)_l^j,
 \nonumber\\[5pt]
 &\hspace{4.cm}\textrm{which proves the relation (\ref{rel12});}\nonumber
 \end{array}\\[6pt]
 \begin{array}{rl}
dv\otimes d^2x^j=&
%
dx^i\otimes d^2x^k\,\xi(v_i)_k^{j}+ q^2\,x^i\,d^2x^k\otimes
d\xi(v_i)_k^j\nonumber\\[3pt]
=&q^2\,d^2x^l\otimes d\xi_l^{ik}\,\xi(v_i)_k^{j}+q^2\,d^2x^l\otimes
\xi_l^{i k}\,d\xi(v_i)_k^j
%
=q^2d^2x^l\otimes d\xi(v)_l^j,\nonumber\\[5pt]
&\hspace{5.cm}\textrm{which proves the relation
(\ref{rel-2-22})};\nonumber
 \end{array}\\[6pt]
 \begin{array}{rl}
d^2(v)\otimes dx^j=&
%
 d^2x^i\otimes dx^k\,\xi(v_i)_k^j
   + q\,[2]_q\,dx^i\otimes dx^k\otimes d\xi(v_i)_k^j\nonumber\\
 & \hspace{2.1cm}+\,(q-1)\,x^id^2x^k\otimes
d\xi(v_i)_k^{j}+ q^2\,x^i\,dx^k\otimes d^2\xi(v_i)_k^j\nonumber\\[3pt]
%
%
=&(q-1)\,d^2x^l\otimes\big(d\xi_l^{ik}\,\xi(v_i)_k^j+
\xi_l^{ik}\,d\xi(v_i)_k^j\big)\nonumber\\[3pt]
&\hspace{0.1cm} +\,q^2\,dx^l\otimes\big(d^2\xi_l^{ik}\otimes
\xi(v_i)_k^j+ {[2]}_{q}\,d\xi_l^{ik}\otimes
d\xi(v_i)_k^j+\xi_l^{ik}d^2\xi(v_i)_k^j\big)\nonumber\\[3pt]
=& (q-1)\,d^2x^l\otimes d\xi(v)_l^j+ q^2\,dx^l\otimes
d^2\xi(v)_l^j,\nonumber\\[5pt]
&\hspace{5.1cm}\textrm{which proves the relation (\ref{rel22});}
\end{array}\\[6pt]
 \begin{array}{rl}
d^3\big( \xi(v)_k^j \big)=&
d^3\xi_l^{ij}\,\xi(v_i)_k^l +{[3]}_q \,d(xi_l^{ij})\otimes
d^2(\xi(v_i)_k^l)
\nonumber\\[3pt]
&+\,+ {[3]}_q\,d(\xi_l^{ij})\otimes d^2(\xi(v_i)_k^l)+
\xi_l^{ij}\,d^3(\xi(v_i)_k^l) \equiv 0,
\nonumber\\[5pt]
 &\hspace{5.3cm}\textrm{which proves the
relation (\ref{rel32});}\nonumber
\end{array}\\[6pt]
 \begin{array}{rl}
d^2(v)\otimes d^2x^j=&
d^2x^i\otimes d^2x^k\,\xi(v_i)_k^{j}+q^2[2]_q\,dx^i\otimes
d^2x^k\otimes
d\xi(v_i)_k^j\nonumber\\[3pt]
 &\hspace{3.0cm}+\,q\,x^i\,d^2x^k\otimes d\xi(v_i)_k^j\nonumber\\
=& q\,d^2x^l\otimes \big(d^2\xi_l^{i k}\,\xi(v_i)_k^{j}+
{[2]}_q\,d\xi_l^{ik}\otimes d\xi(v_i)_k^j+\xi_l^{ik}\,d^2\xi(v_i)_k^j\big)\nonumber\\[3pt]
 =&
q\,d^2x^l\otimes
d^2\xi(v)_l^j,\nonumber\\[-3pt]
&\hspace{5.0cm}\textrm{which proves the relation
(\ref{rel42})}\nonumber
 \end{array}
 \eeq
 and completes the proof of the proposition.


\section{Proof of Theorem \ref{mainthm}}

In order to prove Theorem \ref{mainthm}, we need to show the
following properties of the ideal $I_q$ in
$T_{\mcA}\mE=\bigoplus_{n\geq 0}T^n$:
\begin{enumerate}
\item \label{proof1} $d(I_q)\subset I_q$\ (i.e. the ideal $I_q$ is
$d$-compatible);\\[-8pt]~
 \item \label{proof2}
 the $q$-Leibniz rule
 \beq \label{L-rule3}
  d(\omega
\theta)= (d\omega )\theta + q^{n}\omega (d\theta)\ \ (mod\ I_q)
 \eeq
is satisfied $\forall\ \omega \in T^{n}$ and $ \theta \in
T_{\mcA}\mE$;\\[-6pt]~
 \item \label{proof3}
  $d^3(T_{\mcA} \mE))\subseteq I_q$.
\end{enumerate}
We remind the reader that all tensor products are taken over the
algebra $\mcA$.

 We begin by proving the property (\ref{proof2}), which we do by induction
on the degree of the form $\theta \in T_{\mcA}\mE$. We first show
that the $q$-Leibniz rule (\ref{L-rule3}) holds for the elements:
$v\in\mcA $,\ \ ${\theta}_1 = v_k\,dx^k \in \mcM \equiv T^1$,\ \
and\ \ ${\theta}_2 = v_k\,d^2x^k \in \Msec$,\ \ where $v_k \in \mcA
$.

Let $\omega = d^{a_1}x^{i_1}\dots d^{a_m}x^{i_m}\,r_{i_1\dots i_m}
\in T^{n}$, i.e. $a_1+\dots +a_m=n$.

 Using  the definition of operator $d$ (\ref{d}) and the relations
 from Proposition \ref{prop}, we get
 \beq
 \begin{array}{rl}
 d(\omega\, v)=& \!\!
 \sum_{j=1}^m \,q^{a_1+\dots +a_j}d^{a_1}x^{i_1}\otimes\dots
 \otimes\ d^{a_{j-1}}x^{i_{j-1}}\otimes d^{a_{j}+1}x^{i_{j}}\otimes
 d^{a_{j+1}}x^{i_{j+1}}\otimes
 \nonumber\\[2pt]
&\hspace{6.8cm} \dots\otimes d^{a_m}x^{i_m}\,r_{i_1\dots i_m}\,v
 \nonumber\\[3pt]
&+\, q^{a_1+\dots +a_m}\, d^{a_1}x^{i_1}\otimes \dots \otimes
d^{a_m}x^{i_m}\otimes d(r_{i_1\dots
 i_m})\,v\nonumber\\[3pt]
 &+\,q^{a_1+\dots +a_m}\, d^{a_1}x^{i_1}\otimes \dots \otimes
d^{a_m}x^{i_m}\,r_{i_1\dots i_m}\otimes d(v)
\\[3pt]
&\hspace{-.6cm}=\, d(\omega)\,v+q^{n}\, \omega\otimes dv,\nonumber
 \end{array}\nonumber \\[6pt]
%
 \begin{array}{rl}
 d(\omega \otimes\theta_1)=& \!\!
 d\big(d^{a_1}x^{i_1}\otimes \dots \otimes d^{a_m}x^{i_m}\otimes
 dx^l \xi(r_{i_1\dots i_m})_s^k \xi(v_k)_l^s\big)\nonumber\\[3pt]
 =& \!\! \big(\sum_{j=1}^m \,q^{a_1+\dots +a_j}
 d^{a_1}x^{i_1}\otimes \dots \otimes
 d^{a_{j-1}}x^{i_{j-1}}\otimes d^{a_{j}+1}x^{i_{j}}\otimes
 d^{a_{j+1}}x^{i_{j+1}}\otimes
  \nonumber\\[2pt]
&
 \hspace{7.3cm}\dots\otimes d^{a_m}x^{i_m}\,r_{i_1\dots i_m}  \nonumber\\[3pt]
&+\, q^{a_1+\dots +a_m}\, d^{a_1}x^{i_1}\otimes\dots\otimes
  d^{a_m}x^{i_m}\otimes d(r_{i_1\dots i_m})\big)\otimes v_kdx^k\nonumber\\[3pt]
 &+\,q^{a_1+\dots +a_m}\, d^{a_1}x^{i_1}\otimes\dots\otimes d^{a_m}x^{i_m}\,r_{i_1\dots
 i_m}\otimes\big(dv_k\otimes dx^k+
 v_kd^2x^k\big)\nonumber\\[3pt]
 =& \!\!d(\omega)\otimes\theta_1+q^n\,\omega\otimes d(\theta_1);\nonumber
 \end{array}\\[6pt]
 \begin{array}{rl}
 d(\omega \otimes \theta_2)=&\!\! 
 d\big(d^{a_1}x^{i_1}\otimes\dots\otimes d^{a_m}x^{i_m}
 \otimes d^2x^l \xi(r_{i_1\dots i_m})_s^k\xi(v_k)_l^s\big)\nonumber\\[3pt]
 =&\!\!
 \big(\sum_{j=1}^m \,q^{a_1+\dots +a_j}\!
 d^{a_1}x^{i_1}\otimes\dots \otimes
 d^{a_{j-1}}x^{i_{j-1}}\otimes d^{a_{j}+1}x^{i_{j}}\otimes
 d^{a_{j+1}}x^{i_{j+1}}\otimes\nonumber\\[2pt]
 & \hspace{7.1cm}\dots
 \otimes d^{a_m}x^{i_m}\,r_{i_1\dots i_m} \nonumber\\[3pt]
 &+\, q^{a_1+\dots +a_m}\otimes d^{a_1}x^{i_1}\otimes\dots
  \otimes
  d^{a_m}x^{i_m}r_{i_1\dots i_m}\big)\otimes
  v_kd^2x^k \nonumber\\[3pt]
 &+\,q^{a_1+\dots +a_m}\, d^{a_1}x^{i_1}\otimes\dots
 \otimes d^{a_m}x^{i_m}\,r_{i_1\dots
 i_m}\otimes dv_k\otimes d^2x^k\nonumber\\[3pt]
 =&\!\! d(\omega)\otimes \theta_2+q^n\,\omega\otimes d(\theta_2),\nonumber
\end{array}
 \eeq
as desired.

Assume the $q$-Leibniz rule (\ref{L-rule3}) holds for the form $\rho
\in T^l$; we will then prove it for the forms $\theta = \rho\otimes
dx^k\,r_k\in T^{l+1}$, $r_k \in\mcA$.
 We have
 \beq
 \begin{array}{rl}
 d(\omega \otimes \theta)=& 
d(\omega\otimes \rho)\otimes dx^k\,r_k +
q^{n+l}\omega\otimes\rho\otimes d(dx^k\,r_k)\nonumber\\[3pt]
=&d(\omega)\otimes \rho\otimes dx^k\,r_k + q^n\omega\otimes
d(\rho)\otimes dx^k\,r_k+ q^{n+l}\omega\otimes\rho\otimes
d(dx^k\,r_k) \nonumber\\[3pt]
=&
d(\omega)\otimes \theta + q^n\omega\otimes d(\theta). \nonumber
\end{array}
 \eeq
Analogously for $\theta=\rho\otimes d^2x^k\,r_k \in T^{l+2}$, we
obtain
 \beq
 \begin{array}{rl}
 d(\omega \otimes \theta)=&
 d(\omega\otimes\rho)
 \otimes d^2x^k\,r_k +q^{n+l}\omega\otimes\rho
 \otimes d(d^2x^k\,r_k)
 \nonumber\\[3pt]
=&d(\omega)\otimes\rho
 \otimes d^2x^k\,r_k +q^{n}\omega\otimes d(\rho)
 \otimes d^2x^k\,r_k+q^{n+l}\omega\otimes\rho
 \otimes d(d^2x^k\,r_k)
 \nonumber\\[3pt]
=&
d(\omega)\otimes \theta+q^n\,\omega\otimes d(\theta).\nonumber
 \end{array}
 \eeq
The property (\ref{proof2}) is proved. $\square$


Further, in order to prove the properties (\ref{proof1}) and
(\ref{proof3}), we need in the following
\begin{lemma}\label{d3A}
$d^3(\mcA)\subseteq I_q$.
\end{lemma}
\textbf{Proof.\quad} Since $d^3x^i=0$, we can prove the lemma by
induction.

Let $v=x^i\,v_i$, where $d^3(v_i)\in I_q$ is assumed. Since the
assertion (\ref{proof2}) is true, a straightforward calculation
shows that
 \beq
 \begin{array}{rl}
d^3(v)=
d^3x^{i}\,v_i+ {[3]}_q \,d^2x^{i}\otimes
dv_i + {[3]}_q\,dx^{i}\otimes d^2v_i+x^{i}\,d^3v_i
=x^{i}\,d^3v_i \in x^iI_q \subseteq I_q, \nonumber
 \end{array}
 \eeq
 which is the desired inclusion. $\square$

Next we prove the property (\ref{proof1}).

 Since the generators of $I_q$ (\ref{rel1})\,--\,(\ref{rel4}) were
 obtained from the basic relations (\ref{basic}) by consecutive differentiation,
 we need to check only that the differentials of the two last
 generators (\ref{rel3}) and (\ref{rel4})
 belong to the ideal $I_q$. By direct calculation, we get
\beq
 \begin{array}{rl}
&\hspace{-.3cm}
  d(d^3\xi_k^{ij}) \nonumber\\[3pt]    
& = {[2]}_q\,d^2x^l\otimes d^2x^mD_mD_l(\xi_k^{ij})+
q\,[2]_q\,d^2x^l\otimes dx^m\otimes dx^pD_pD_mD_l(\xi_k^{ij})\nonumber\\[3pt]
&\ \ +\, q^2d^2x^l\otimes d^2x^mD_mD_l(\xi_k^{ij})+
q^2dx^l\otimes d^2x^m\otimes dx^pD_pD_mD_l(\xi_k^{ij})\nonumber\\[3pt]
&\ \ +\, d^2x^l\otimes dx^m\otimes dx^pD_pD_mD_l(\xi_k^{ij})+
qdx^l\otimes d^2x^m\otimes dx^pD_pD_mD_l(\xi_k^{ij})\nonumber\\[3pt]
&\ \ +\, q^2dx^l\,{\otimes}\, dx^m\,{\otimes}\,
d^2x^pD_pD_mD_l(\xi_k^{ij})+
dx^l\,{\otimes}\, dx^m\,{\otimes}\, dx^pD_p\,{\otimes}\,
dx^sD_s\,D_m\,D_l(\xi_k^{ij})\nonumber\\[5pt]
 &= [3]_q\,d^2x^l\otimes
d^2x^mD_mD_l(\xi_k^{ij})+
[3]_q\,d^2x^l\otimes dx^m\otimes dx^pD_pD_mD_l(\xi_k^{ij})\nonumber\\[3pt]
&\ \ +\, q[2]_q\,dx^l\otimes d^2x^m\otimes
dx^pD_pD_mD_l(\xi_k^{ij})+
q^2\,dx^l\otimes dx^m\otimes d^2x^pD_p\,D_m\,D_l(\xi_k^{ij})
\nonumber\\[3pt]
&\ \ +\, dx^l\otimes dx^m\otimes dx^p\otimes
dx^sD_sD_p\,D_m\,D_l(\xi_k^{ij})= dx^l\otimes d^3(D_l(\xi_k^{ij})).
 \end{array}
 \eeq
Recall that $D_l(\xi_k^{ij})\in \mcA $.  Lemma \ref{d3A} implies the
inclusion $d^3(D_l(\xi_k^{ij}))\in I_q $, from which it follows that
$dx^l\otimes d^3(D_l(\xi_k^{ij}))\in dx^l\otimes I_q \subseteq I_q$.

Analogously,
 \beq
\begin{array}{l}\bigskip
d(d^2x^i\otimes d^2x^j-qd^2x^k\otimes d^2\xi_k^{ij})=
-d^2x^k\otimes d^3\xi_k^{ij}\in d^2x^k\otimes I_q \subseteq I_q,
\nonumber
\end{array}
 \eeq
and hence the property (\ref{proof1}) is proved.

 In order to prove the property (\ref{proof3}), it is sufficient to
 show that $d^3 (T^n)\subseteq I_q$. Since the case $n=0$, i.e.
 $T^0=\mcA$, is already proved (see Lemma \ref{d3A}), this enables us to
prove the desired inclusion by induction on the degree of the forms
from $T_{\mcA} \mE$.

  Let $\omega _1=dx^k\otimes \theta _k$, where $\theta _k \in
  T^{n-1}$, and $\omega _2=d^2x^k\otimes \rho _k$, where $\rho _k \in
  T^{n-2}$, and the property
  (\ref{proof3})  be satisfied for the forms $\theta _k$ and $\rho
  _k$.
  Consequently applying the operator $d$  to the forms
  $\omega _1$ and $\omega _2$, we obtain the following inclusions for $\omega _1$:
  \beq
\begin{array}{rl}
&d(\omega_1)\in d^2x^k\,\theta_k + qdx^k\,d\theta_k + I_q;
\nonumber\\[5pt]
&d^2(\omega_1) \in q[2]_q d^2x^k\,d\theta_k+q^2dx^k\,d^2\theta_k
+d(I_q)+I_q; \nonumber\\[5pt]
&d^3(\omega_1)\in dx^k\,d^3\theta_k+d^2(I_q)+d(I_q)+I_q\subseteq
d^2x^k\, I_q+d^2(I_q)+d(I_q)+I_q\subseteq I_q;
\end{array}
 \eeq
and for $\omega_2$:
 \beq
\begin{array}{rl}
& d(\omega_2)\in q^2d^2x^k\,d\rho_k + I_q;
\nonumber\\[5pt]
& d^2(\omega_2) \in q d^2x^k\,d^2\rho_k+d(I_q)+I_q;
\nonumber\\[5pt]
& d^3(\omega_2)\in d^2x^k\,d^3\rho_k+d^2(I_q)+d(I_q)+I_q \subseteq
d^2x^k\, I_q+d^2(I_q)+d(I_q)+I_q\subseteq I_q.
\end{array}
 \eeq
The proof of the theorem is thus completed. $\square$
\medskip


 \def\comp  {Com\-mun.\,Math.\,Phys.}
 \def\czjp  {Czech.\,J.\,Phys.}
 \def\jgap  {J.\,Geom. and\,Phys.}
 \def\jomp  {J.\,Math.\,Phys.}
 \def\lemp  {Lett.\,Math.\,Phys.}

\end{document}